\documentclass[pre,aps,epsf,preprint,superscriptaddress]{revtex4}
\usepackage{psfig}
\usepackage{amssymb}
\usepackage{graphicx}
\usepackage{graphics}
\usepackage{epsfig}
\newcommand{\be}{\begin{equation}}
\newcommand{\ee}{\end{equation}}

\begin{document}

\title{Space Charges Can Significantly Affect the Dynamics of Accelerator Maps}

\author{Tassos~Bountis}
\affiliation {Department of Mathematics, University of Patras,
GR-26500 Patras, Greece}

\affiliation {Center for Research and Applications of Nonlinear
Systems (CRANS), University of Patras, GR-26500 Patras, Greece}

\author{Charalampos~Skokos}
\affiliation {Center for Research and Applications of Nonlinear
Systems (CRANS), University of Patras, GR-26500 Patras, Greece}

\begin{abstract}

Space charge effects can be very important for the dynamics of
intense particle beams, as they repeatedly pass through nonlinear
focusing elements, aiming to maximize the beam's luminosity
properties in the storage rings of a high energy accelerator. In
the case of hadron beams, whose charge distribution can be
considered as ``frozen" within a cylindrical core of small radius
compared to the beam's dynamical aperture, analytical formulas
have been recently derived \cite{BenTurc} for the contribution of
space charges within first order Hamiltonian perturbation theory.
These formulas involve distribution functions which, in general,
do not lead to expressions that can be evaluated in closed form.
In this paper, we apply this theory to an example of a charge
distribution, whose effect on the dynamics can be derived
explicitly and in closed form, both in the case of 2--dimensional
as well as 4--dimensional mapping models of hadron beams. We find
that, even for very small values of the ``perveance" (strength of
the space charge effect) the long term stability of the dynamics
changes considerably. In the flat beam case, the outer invariant
``tori" surrounding the origin disappear, decreasing the size of
the beam's dynamical aperture, while beyond a certain threshold
the beam is almost entirely lost. Analogous results in mapping
models of beams with 2-dimensional cross section demonstrate that
in that case also, even for weak tune depressions, orbital
diffusion is enhanced and many particles whose motion was bounded
now escape to infinity, indicating that space charges can impose
significant limitations on the beam's luminosity.

\end{abstract}
\maketitle
\section {Introduction}
\label{INTRO}

One of the fundamental problems concerning the dynamics of
particle beams in the storage rings of high energy accelerators is
the determination of the beam's {\it dynamical aperture}, i.e. the
maximal domain containing the particles closest to their ideal
circular path and for the longest possible time. For example,
``flat" hadron beams (experiencing largely horizontal betatron
oscillations) can be described by 2--dimensional (2D)
area--preserving maps, where the existence of invariant curves
around the ideal stable path at the origin, guarantees the
stability of the beam's dynamics for infinitely long times
\cite{BTTS94,GT96}. The reason for this is that the chaotic motion
between these invariant curves is always bounded and the particles
never escape to infinity. Other important phenomena that arise in
this context is the presence of a major resonance in the form of a
chain of islands through which the beam may be collected, or the
existence of an outer invariant curve surrounding these islands,
which serves as a boundary of the motion and thus allows an
estimate of the beam's dynamical aperture.

On the other hand, hadron beams with a 2 - dimensional cross
section require the use of $4$--dimensional (4D) symplectic
mappings for the study of their dynamics
\cite{BT91,BK94,VBK96,VIB97,BS05}. In fact, if longitudinal (or
synchrotron) oscillations also need to be included the mappings
become 6--dimensional. In such cases, the problems of particle
loss are severe, as chaotic regions around different resonances
are connected, providing a network of paths through which
particles can move away from the origin, and eventually escape
from the beam after sufficiently long times.

In this Letter, we add to these issues the presence of space
charges within a core radius $r_c$, which is small compared to the
beam's dynamical aperture. In other words, we will assume that our
proton (or antiproton) beam is intense enough so that the effect
of a charge distribution concentrated within this core radius
cannot be neglected.  Furthermore, we will consider this
distribution as cylindrically symmetric and ``frozen" (i.e. time
independent, so that it may be self consistent with the linear
lattice) and study the dynamics of a hadron beam as it passes
repeatedly through nonlinear magnetic focusing elements of the
FODO cell type. This system has been studied extensively in the
absence of space charge effects in
\cite{BTTS94,GT96,BT91,BK94,VBK96,VIB97,BS05} and the question we
raise now is whether its dynamics can be seriously affected if
space charges are also taken into consideration.

Space charge presents a fundamental limitation to high intensity
circular accelerators. Its effects are especially important in the
latest designs of high-intensity proton rings, which require beam
losses much smaller than presently achieved in existing
facilities. It is therefore necessary to estimate the major space
charge effects which could lead to emittance growth and associated
beam loss \cite{Fedotov}. The interplay between nonlinear effects,
typical of single-particle dynamics, and space charge, typical of
multi-particle dynamics induced by the Coulomb interaction,
represents a difficult challenge. To understand better these
phenomena, an intense experimental campaign was launched at the
CERN Proton Synchrotron \cite{Franchetti_2003}. It is very
important, therefore, to develop analytical techniques which could
be utilized in order to study and localize the associated web of
resonances (see e.g. \cite{PAC2001}) to obtain an analytical
estimation of the dynamic aperture, as suggested e.g. in
\cite{Benedetti}.

In a recent paper, Benedetti and Turchetti \cite{BenTurc} used
first order canonical perturbation theory to obtain analytical
expressions for the jump in the position and momenta due to the
multipolar kicks in such maps, showing that the space charges
effectively modulate the tune at every passage of the particle
through a nonlinear element of the lattice. In particular, they
derived the new position and momentum coordinates after the $n$th
passage through a FODO cell in the thin lens approximation, as the
iterates of the 2D map
\begin{equation}
\left( \begin{array}{c} X_{n+1} \\P_{n+1}
\end{array} \right)  =
\left( \begin{array}{cc}
\cos \Omega(J) & -\sin \Omega(J) \\
\sin \Omega(J) &  \ \cos \Omega(J) \\
\end{array} \right)
 \times  \left( \begin{array}{c}
X_{n} \\
P_{n}+X_{n}^{k-1} \\
\end{array} \right), \ \ n=0,1,2,...,
\label{eq:map}
\end{equation}
where
\begin{equation}
J = \frac{P_n^2+X_n^2}{2}, \label{eq:action}
\end{equation}
for $k=3$, i.e. in the case of sextupole nonlinearities and
\begin{equation}
\Omega(J) = \omega + \frac{\omega_0^2-\omega^2}{2\omega}
\left(1-\frac{R_c^2}{J}
\frac{1}{2\pi}\int_0^{2\pi}g_1\left(\frac{2J\cos^2\theta}{R_c^2}\right)d\theta
\right). \label{eq:Omega}
\end{equation}
The variables $P$, $X$ and the parameters entering in the above
expressions are related to the corresponding ones of the original
Hamiltonian
\begin{equation}
H=\frac{p^2}{2}+\omega_0^2\frac{x^2}{2}-\frac{x^3}{3}\sum_{l=1,2,..}\delta(s-l)-\frac{\xi}{2}
g_2\left( \frac{x^2}{r_c^2}\right), \label{eq:Ham1}
\end{equation}
by the formulas
\begin{equation}
P = p/ \omega^2  \ , \  X=x/\omega \ , \ \ R_c=r_c \omega^{-3/2} \
, \ \ \omega^2 = \omega_0^2-\frac{\xi}{r_c^2}, \label{eq:param}
\end{equation}
where $\omega$ is the depressed phase advance at the center of the
charge distribution, $p=dx/ds$, $s$ is the coordinate along the
ideal circular orbit, $g_2(t)$ is given by
\begin{equation}
g_2(t)=\int_0^tu^{-1}g_1(u)du \ , \ g_1(t)=\int_0^tg(u)du  \ , \
g\left( \frac{r^2}{r_c^2} \right)=\pi r_c^2\rho(r) \label{eq:dens}
\end{equation}
$\rho(r)$ satisfies $\int_0^{\infty}\pi \rho(r) dr^2=1$, $\pi
r_c^2\rho(0)=1$ and $Q\rho(r)$ represents the radial charge density.
Note that if $m$, $q$, and $v_0$ denote the mass, charge and
velocity of our non-relativistic particles, the ``perveance"
parameter $\xi=2qQ/mv_0^2$ determines the tune depression in
(\ref{eq:param}), which must be small for the above analysis to be
valid \cite{BenTurc}.

The stage is now set for the investigation of space charge effects
on the dynamics. However, the space advances $\Omega(J)$ needed in
(\ref{eq:map}) at every iteration depend on integrals of the
distribution function $g(u)$ that may not be available
analytically. To overcome this difficulty, we choose in section II
a particular form of $g(u)$ for which these integrals can be
explicitly carried out not only for 2D maps of the ``flat" beam
case, but also for 4D maps describing vertical as well as
horizontal deflections of the beam's particles.

Thus, in section III we perform numerical experiments to examine
the influence of space charges on the dynamics and find indeed
that even for small perveance values the long term stability of
the beam is significantly affected. In particular, as $\xi$ grows
(or the tune depression $\omega/\omega_0$ decreases from 1),
perturbations of 2D as well as 4D maps show that the outer
invariant ``surfaces" surrounding the ideal circular path at the
origin disappear and the beam's dynamical aperture is seriously
limited. Only the major unperturbed stable resonances survive,
with their ``boundaries" clearly diminished by the presence of new
resonances due to space charge effects. In our 2D mapping model, a
threshold value of $\xi$ (or $\omega/\omega_0$) was found, beyond
which the beam is practically destroyed. The paper ends by
describing our concluding remarks and work in progress in section
IV.

Space charge effects on beam stability became a relevant topic
during the years when construction of medium-low energy high
currents  accelerators, such as SNS and the design of the FAIR
rings at GSI were  started \cite{Jeon,PAC2001,Franchetti_2005}
(see also many articles in the SNS Accelerator Physics Proceedings
of the last few years). The role of collective effects and
resonances has attracted considerable attention, since they can
cause significant beam quality deterioration and losses
\cite{Hofmann}. Another relevant issue is the coupling with the
longitudinal motion which modulates the transverse tune and
induces losses by resonance crossing as shown by recent
experiments \cite{Franchetti_2003}.

High intensity rings, where the bunches can circulate over one
million turns, require a careful analysis of the long term
stability of the beam. Since the commonly used codes require large
CPU times and exhibit an emittance growth due to numerical noise,
they are not suited for long term dynamic aperture studies and the
use of faster methods is necessary \cite{Franchetti_2005}. The
method proposed in \cite{BenTurc} allows us to introduce space
charge effects in one single evaluation of the map, when a thin
sextupole or octupole is present, just as one does in the absence
of space charge, and is thus especially well suited for dynamical
aperture calculations.

\section {Exact Results for a Specific Charge Distribution}
\label{Exact}

\subsection {The One - dimensional Beam}
\label{One_Dim}

Let us choose for our space charge distribution function in the
1--dimensional case the form
\begin{equation}
g\left( \frac{X^2}{R_c^2}\right)= \frac{1}{(X^2/R_c^2+1)^2}\,\,.
\label{eq:distr1}
\end{equation}
The generalization to 2 dimensions is evident by replacing $X^2$
by $X^2+Y^2$ in (\ref{eq:distr1}). Observe that this function
satisfies the requirements that $g(0)=1$, $g_1(t)\propto t$ as
$t\rightarrow 0$ and $g_1(t)\rightarrow 1$ as $t\rightarrow 0$,
(using (\ref{eq:dens})), as expected from the theory
\cite{BenTurc}.

Evaluating now by elementary manipulations the integral in
(\ref{eq:Omega}), using (\ref{eq:distr1}) and (\ref{eq:dens}), we
find that it is given by the closed form expression
\begin{equation}
\int_0^{2\pi}g_1\left(\frac{2J\cos^2\theta}{R_c^2}\right)
d\theta=2\pi-\frac{2\pi}{(\frac{2J}{R_c^2}+1)^{1/2}}\,\,.
\label{eq:int}
\end{equation}
Thus, the phase advance at every iteration becomes
\begin{equation}
\Omega(J) = \omega + \frac{\omega_0^2-\omega^2}{2\omega}
\left(1-\frac{R_c^2}{J}+ \frac{R_c^2}{J(\frac{2J}{R_c^2}+1)^{1/2}}
\right), \label{eq:Omega1}
\end{equation}
where $J$ is given by (\ref{eq:action}). Note that, in the limit
$J\rightarrow 0$, eq.~(\ref{eq:Omega1}) implies that
$\Omega\rightarrow \omega$ as expected. In fact, expanding the
square root in that limit we find
\begin{equation}
\Omega(J) = \omega + \frac{\omega_0^2-\omega^2}{2\omega}
\left(\frac{3}{2}\frac{J}{R_c^2}- \frac{5}{2}\frac{J^2}{R_c^4}+...
\right), \label{eq:Omega1b}
\end{equation}
from which we can estimate the deviation of $\Omega$ from the
depressed tune $\omega$ near the origin. In section III below we
pick an $\omega_0$ such that for $\xi=0$ we have a major resonance
and a relatively large dynamical aperture in the $X_n, P_n$ plane,
select $r_c$ small compared with this aperture and vary $\xi$ to
study the space charge effect on the dynamics.

\subsection {The Two - dimensional Beam}
\label{Two_Dim}

Let us now observe that in two space dimensions the original
Hamiltonian of the system, (\ref{eq:Ham1}), becomes
\begin{equation}
H=\frac{p_1^2}{2}+\omega_{01}^2\frac{x^2}{2}+\frac{p_2^2}{2}+
\omega_{02}^2\frac{y^2}{2}+\left(-\frac{x^3}{3}+xy^2\right)
\sum_{l=1,2,..}\delta(s-l)-\frac{\xi}{2}
g_2\left(\frac{x^2+y^2}{r_c^2}\right), \label{eq:Ham2}
\end{equation}
where sextupole nonlinearities involve, of course, both $x$ and
$y$ variables. Since there are now two tune depressions
\begin{equation}
\omega_1=\left(\omega_{01}^2-\frac{\xi}{r_c^2}\right)^{1/2} \  \ ,
\ \ \omega_2=\left(\omega_{02}^2-\frac{\xi}{r_c^2}\right)^{1/2},
\label{eq:omegas}
\end{equation}
after transforming to new variables $X=x \omega_1^{1/2}$, $P_1=p_1
\omega_1^{-1/2}$ and $Y=y \omega_2^{1/2}$, $P_2=p_2
\omega_2^{-1/2}$ defined by
\begin{equation}
X=(2J_1)^{1/2} \cos \theta_1 \ , \ P_1=-(2J_1)^{1/2} \sin \theta_1
\ , \ Y=(2J_2)^{1/2}\cos \theta_2 \ , \ P_2=-(2J_2)^{1/2}\sin
\theta_2 , \label{eq:newvar}
\end{equation}
differentiating the Hamiltonian with respect to $J_1$ and $J_2$
and integrating over $\theta_1$ and $\theta_2$, we find the two
tune depressions
\begin{equation}
\Omega_1 = \omega_1 + \frac{\omega_{01}^2-\omega_1^2}{2\omega_1}
\left(1- \frac{1}{(2\pi)^2}
\int_0^{2\pi}\int_0^{2\pi}\frac{2\cos^2\theta_1}{A+1}d\theta_1d\theta_2
\right) \label{eq:Omega2}
\end{equation}
and $\Omega_2$, with 1 replaced by 2 in (\ref{eq:Omega2}), while A
is defined by
\begin{equation}
A =
\frac{2J_1\cos^2\theta_1}{r_1^2}+\frac{2J_2\cos^2\theta_2}{r_2^2}
\ , \ \label{eq:A}
\end{equation}
where $r_1=r_c \omega_1^{1/2}$, $r_2=r_c \omega_2^{1/2}$.

Observe that we have used for the $g_1$ function under the
integral sign (see (\ref{eq:Omega})), the expression
$g_1(A)=A/(1+A)$, derived from our simple choice of the
distribution function (\ref{eq:distr1}) using (\ref{eq:dens}).

The above $\Omega_1$ and $\Omega_2$ are to be used in the
iterations of the 4D mapping:
\begin{eqnarray}
\left( \begin{array}{c} X(n+1) \\P_1(n+1) \\Y(n+1) \\P_2(n+1)
\end{array} \right) & = &
\left( \begin{array}{cccc}
\cos \Omega_1 & -\sin \Omega_1 & 0 & 0 \\
\sin \Omega_1 & \cos \Omega_1 & 0 & 0 \\
0 & 0 & \cos \Omega_2 & -\sin \Omega_2 \\
0 & 0 & \sin \Omega_2 & \cos \Omega_2
\end{array} \right) \nonumber \\
 & \times & \left( \begin{array}{c}
X(n) \\
P_1(n)+X^2(n)-Y^2(n) \\
Y(n) \\
P_2(n)-2 X(n) Y(n)
\end{array} \right),
\label{eq:map2}
\end{eqnarray}
whose dynamics has already been studied extensively in
\cite{BT91,BK94,VBK96,VIB97,BS05} in the absence of space charge
effects, i.e for $\omega_1=\omega_{01}$ and
$\omega_2=\omega_{02}$. In these papers, it was observed that for
the tune values $q_x=0.61903$, and $q_y=0.4152$, in
\begin{equation}
\omega_{01} = 2 \pi q_x , \,\,\omega_{02} = 2 \pi q_y,
\label{eq:tunes}
\end{equation}
a large dynamical aperture is achieved, with interesting chains of
resonant ``tori" surrounding the origin. In section III we will
study what happens to these structures when $\xi > 0$ (i.e
$\omega_1 <\omega_{01}$, $\omega_2 <\omega_{02}$) and space charge
effects are taken into account. Before doing this, however, it is
necessary to describe how the integrals in (\ref{eq:Omega2}) are
to be evaluated: Let us first perform the integration over
$\theta_2$, writing
\begin{equation}
K_1=\int_0^{2\pi}\int_0^{2\pi}\frac{2\cos^2\theta_1}{A+1}d\theta_1d\theta_2
= 2\int_0^{2\pi}d\theta_1\cos^2\theta_1I(\theta_1),
\label{eq:IntK}
\end{equation}
where
\begin{equation}
I(\theta_1)=\int_0^{2\pi}\frac{d\theta_2}{2A_1\cos^2\theta_2+B_1}
\ , \ A_1=\frac{J_2}{r_2^2} \ , \
B_1=1+\frac{2J_1\cos^2\theta_1}{r_1^2}\,\,. \label{eq:IntI1}
\end{equation}
The integral (\ref{eq:IntI1}) can be evaluated as before with
elementary functions to yield
\begin{equation}
I(\theta_1)=\frac{2\pi}{[B_1(B_1+2A_1)]^{1/2}} = 2\pi\left[
\left(1+\frac{2J_1\cos^2\theta_1}{r_1^2}\right)
\left(1+\frac{2J_1\cos^2\theta_1}{r_1^2}+
\frac{2J_2}{r_2^2}\right)\right]^{-1/2}\,\, . \label{eq:IntI2}
\end{equation}
Inserting now expression (\ref{eq:IntI2}) into the integral
(\ref{eq:IntK}), changing integration variable to
$\phi=2\theta_1$, we easily arrive, after some simple
manipulations, to the expression
\begin{equation}
K_1=\frac{2\pi r_1^2}{J_1}\int_0^{2\pi}d\phi\frac{cos\phi+1}
{\left[(cos\phi+C_1)(cos\phi+D_1)\right)]^{1/2}}\,\, ,
\label{eq:IntK2}
\end{equation}
where
\begin{equation}
C_1=1+\frac{r_1^2}{J_1} \ , \
D_1=1+\frac{r_1^2}{J_1}\left(1+\frac{2J_2}{r_2^2}\right).
\label{eq:CD}
\end{equation}
We finally make the substitution $u=\tan(\phi/2)$ and rewrite the
above integral in the form
\begin{equation}
K_1=\frac{16\pi r_1^2}{J_1}\int_0^{\infty} \frac{du}{(u^2+1)
\left\{ \left[(C_1-1)u^2+1+C_1\right]
\left[(D_1-1)u^2+1+D_1\right] \right\}^{1/2}}\,\, .
\label{eq:IntK1}
\end{equation}
This is clearly not an elementary integral. Notice, however, that
all terms in the denominator of (\ref{eq:IntK1}) are positive and
as $u\rightarrow\infty$ the integrand vanishes as $u^{-4}$. It is,
therefore, expected to converge very rapidly and may be computed,
at every iteration of the map, using standard routines. For
practical purposes, however, in section III below, we prefer to
compute instead its equivalent form (\ref{eq:IntK2}). Of course,
as explained above, a similar integral, $K_2$, also needs to be
computed (with $1\rightarrow2$ in (\ref{eq:IntK}),
(\ref{eq:IntK2}) and (\ref{eq:CD})), whence $\Omega_1$ and
$\Omega_2$ are found and the next iteration of the 4D map
(\ref{eq:map2}) can be evaluated.

\section{Numerical Results}

\begin{figure}[ht]
\begin{center}
\includegraphics{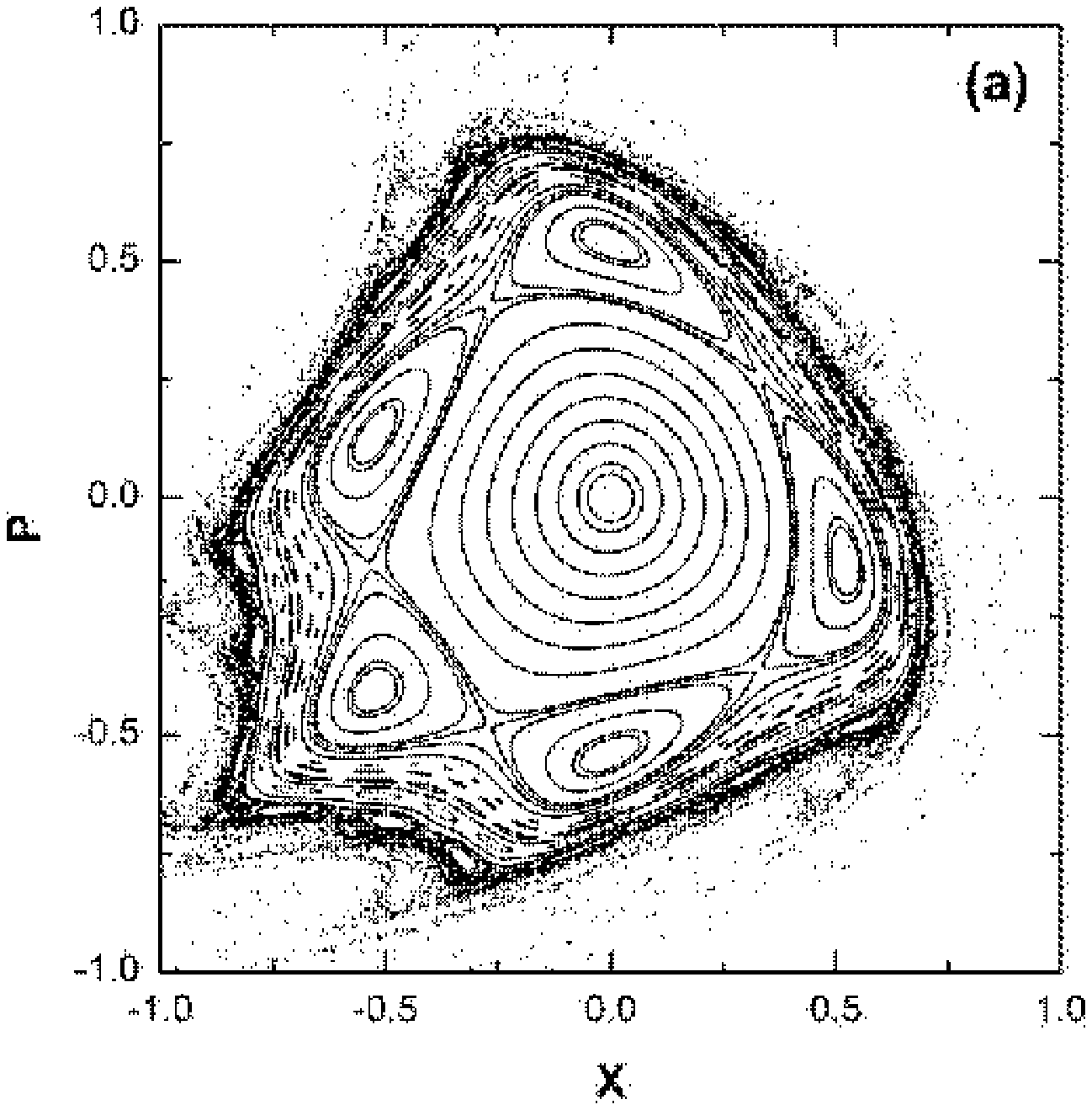} \includegraphics{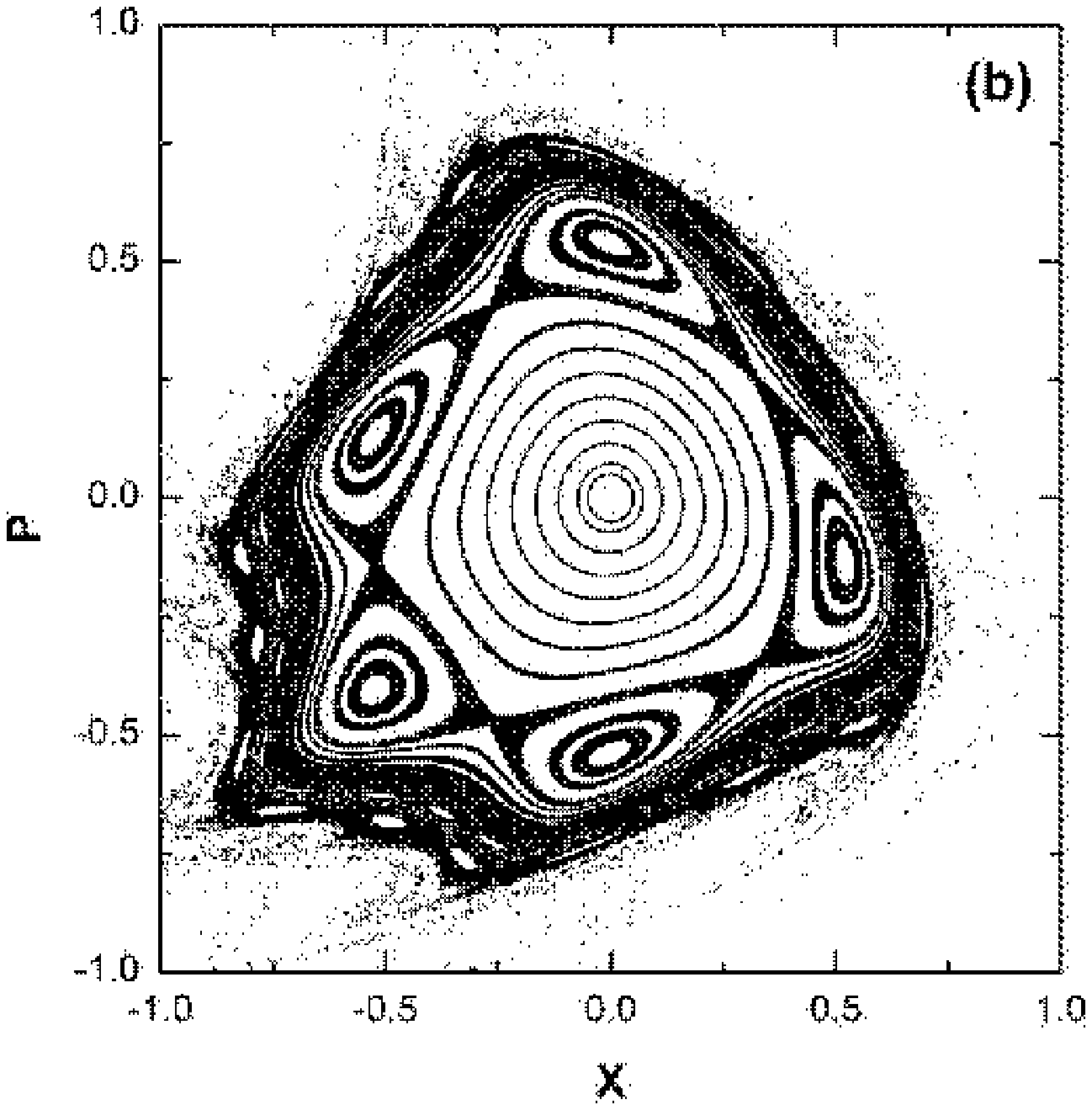} \vspace{8cm}
\end{center}
\caption{(a) The phase space picture of a flat beam, resulting
from the iterations of the map (\ref{eq:map}), with tune
$q_x=0.21$ and frequency $\omega_0=1.32$, in the absence of space
charges. (b) The same picture with $\xi=0.001$ (or
$\omega/\omega_0 \simeq 0.9426$) in the Hamiltonian
(\ref{eq:Ham1}). Notice the dissolution of the outer invariant
curves and the disappearance of many of the smaller islands of (a)
away from the origin, leading to a significant decrease of the
beam's dynamical aperture. Each initial condition is followed for
$N=10^4$ iterations.} \label{graph_1}
\end{figure}

Let us now turn to some practical applications of the above theory
to specific problems concerning the stability of hadron beams
passing through FODO cell magnetic focusing elements and
experiencing sextupole nonlinearities, as described in
\cite{BTTS94,GT96,BT91}. First, we shall consider the flat beam
case (\ref{eq:map}), for the specific tune value $q_x=0.21$
corresponding to frequency $\omega_0=2\pi q_x=1.32$, exhibiting,
in the absence of space charge perturbations, the phase space
picture shown in Figure \ref{graph_1}(a) below. As we see in this
figure, the region of bounded particle motion extends to a radius
of about 0.54 units from the origin. There are also 5 islands of a
major resonance surrounded by invariant curves (or 1D ``tori"),
whose outermost boundary delimits the so - called dynamical
aperture of the beam. Outside that domain there are chains of
smaller islands (representing higher order resonances) and chaotic
regions through which particles eventually escape to infinity.
This escape occurs, of course, at different speeds due to the well
- known phenomenon of ``stickiness", depending on how close the
orbits are to the invariant curves surrounding the islands.

Let us now consider a space charge distribution of the form
(\ref{eq:distr1}) with a ``frozen core" of radius $r_c=0.1$, which
is small compared with the radius of the beam's dynamical
aperture. Our purpose is to vary the value of the preveance
$\xi>0$, cf. (\ref{eq:Ham1}), starting from $\xi=0$, to estimate
the effects of space charge on the dynamics.

\begin{figure}[ht]
\begin{center}
\includegraphics{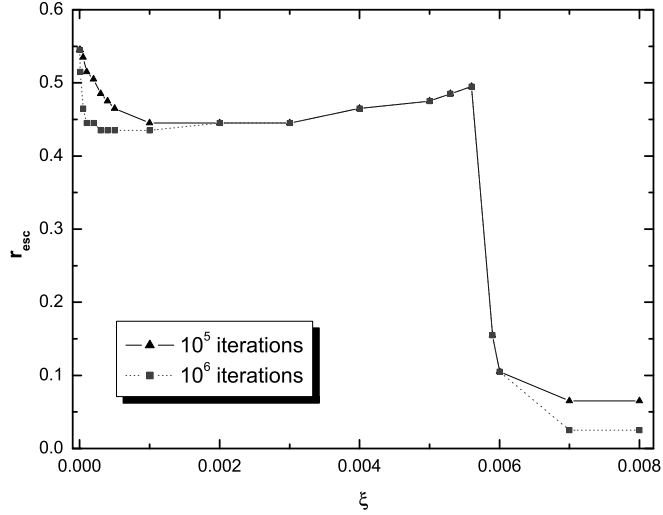}  \vspace{6.5cm}
\end{center}
\caption{Plot of dynamical aperture ``radius" $r_{esc}$ estimates
of the flat beam case, resulting from the iterations of the map
(\ref{eq:map}), with frequency $\omega_0=1.32$, in the presence of
space charges, i.e for increasing values of $\xi$ in
(\ref{eq:Ham1}). Note the quick decrease of the aperture value by
about $20$ percent, which remains nearly the same until $\xi\simeq
0.006$ ($\omega/\omega_0 \simeq 0.81$), when a sudden drop occurs,
probably due to the appearance of a major new resonance.}
\label{graph_2}
\end{figure}

Setting $\xi=0.001$ ($\omega/\omega_0 \simeq 0.97$), for example,
which is quite small compared with $r_c^2=0.01$, we observe in
Figure \ref{graph_1}(b) that the picture has significantly
changed. In particular, the 3-dimensional character of the
dynamics (due to the variation of the space advance $\Omega(J)$)
has turned the invariant curves into ``surfaces" and has led to
the dissolution of the outer ones surrounding the five major
islands. Furthermore, most of the chains of smaller islands of
Figure \ref{graph_1}(a) have disappeared due to the new resonances
caused by the presence of space charges. To see how all this
affects the dynamical aperture of the beam as a function of the
tune depression $\omega/\omega_0$ we now perform the following
experiment:

Forming a grid of initial conditions of step size $\Delta x=\Delta
p=0.01$ within a square $[-1,1]\times[-1,1]$ about the origin
($X_n=P_n=0$), we use (\ref{eq:map}) to iterate for different $\xi
> 0$ (or $\omega/\omega_0 <1$) all points falling within circular rings of
width $\Delta r=0.01$ for $N=10^5$ and $N=10^6$ iterations and
plot in Figure \ref{graph_2} the last
$r=(X_n^2+P_n^2)^{1/2}=r_{esc}$ value, at which at least one orbit
was found to escape from the next outer ring. The results
demonstrate that already at $\xi=0.001$ ($\omega/\omega_0 \simeq
0.97$) our estimate of the dynamical aperture $r_{esc}$ has fallen
from $0.54$ to about $0.43$. In fact, it remains close to that
value (rising somewhat to about $0.5$) until $\xi\simeq .006$
($\omega/\omega_0 \simeq 0.81$), where it experiences a sudden
drop to $r_{esc}\simeq 0.03$, and the beam is effectively
destroyed. Of course, once one orbit escapes most of them quickly
follow within the next one or two circular rings. Note also that
increasing the number of iterations from $N=10^5$ to $N=10^6$ does
not appreciably change the results, until the sudden drop occurs.

This dramatic change at $\omega/\omega_0 \simeq 0.81$ is most
probably due to the presence of a major new resonance caused by
the space charge perturbation. It may be an important effect,
however, since it occurs at a $\xi$ value which is still smaller
than the $r_c^2=0.01$ radius of the charge core. Of course, long
before this happens, already at $\xi\simeq 0.0002$ (or
$\omega/\omega_0 \simeq 0.994$), the effective aperture of the
beam has been significantly reduced by about 20 percent from its
value at $\xi=0$.

\begin{figure}[ht]
\begin{center}
\includegraphics{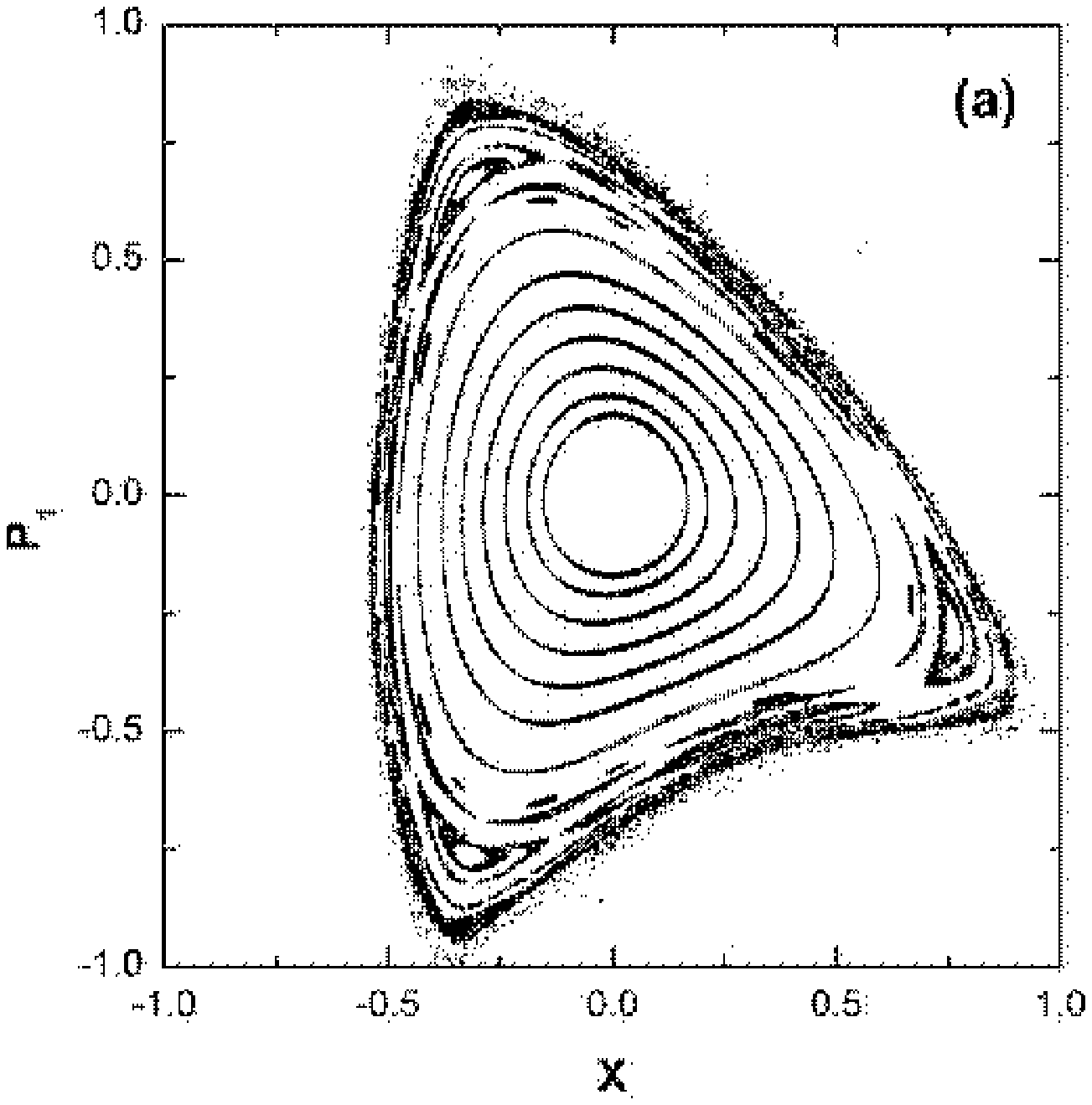} \includegraphics{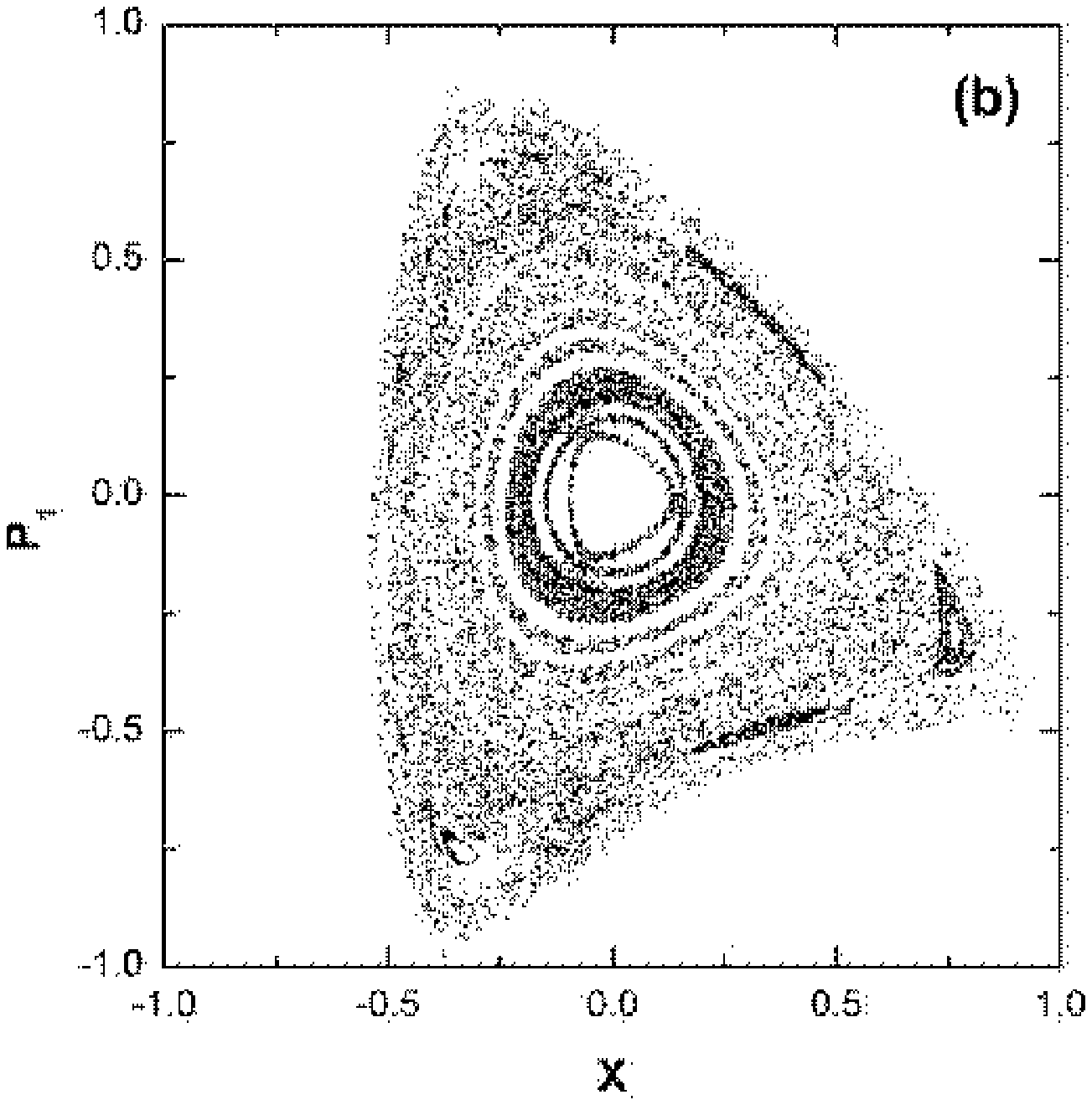} \vspace{8cm}
\end{center}
\caption{(a) An $X,P_1$ projection of a beam with $x$ and $y$
deflections, resulting from the iterations of the map
(\ref{eq:map2}), with tunes $q_x=0.61903, q_y=0.4152$, in the
absence of space charges. Note the presence of a large region of
invariant 2D ``tori" about the origin and the 6 ``islands" of a
stable resonance, shown here in a cut of the 4D space, with
$|Y(n)|\leq 0.04$. (b) The same picture with $\xi=0.0002$ in
Hamiltonian (\ref{eq:Ham2}) (or $\omega_1/\omega_{01} \simeq
0.9993, \omega_2/\omega_{02} \simeq 0.9985$). Each initial
condition is followed for $N=10^4$ iterations. Notice the
dissolution of outer invariant curves surrounding the origin and
the disappearance of the chain of islands of (a), leading to a
significant decrease of the beam's dynamical aperture.}
\label{graph_3}
\end{figure}

Finally, let us turn to the case of the 4D map (\ref{eq:map2}),
describing the more realistic case of a beam whose particles
experience horizontal as well as vertical displacements from the
ideal path, see (\ref{eq:Ham2}). For comparison purposes, we
choose the same parameter values as in our earlier papers
\cite{BK94,VBK96,VIB97,BS05}, i.e horizontal and vertical tunes
$q_x=0.61903$, $q_y=0.4152$ respectively, yielding the unperturbed
frequencies (\ref{eq:tunes}) used in the mapping equations. In
Figure \ref{graph_3}(a), we iterate many initial conditions
$X(0),P_1(0),Y(0),P_2(0)$ around the origin and plot on a
$X(n),P_1(n)$ projection a picture of the dynamics, for
$|Y(n)|\leq 0.04$, in the absence of space charges, i.e. with
$\omega_{i}=\omega_{0i}$, $i=1,2$.

Note the region of invariant tori and a chain of 6 ``islands"
corresponding to a stable resonance. Strictly speaking, the motion
between these tori need not be bounded as 2D surfaces do not
separate 4D space and Arno'ld diffusion phenomena
\cite{Licht-Lieb} could in principle carry orbits far away from
the origin. However, as has been explicitly shown for this model
in \cite{BK94,VBK96,VIB97}, such phenomena are extremely slow and
hence a domain with radius of the order of 0.5 can be effectively
considered as the dynamical aperture of the beam. Repeating this
experiment in the presence of space charges, i.e. with
$\xi=0.0002$ (or $\omega_1/\omega_{01} \simeq 0.9993,
\omega_2/\omega_{02} \simeq 0.9985$) in (\ref{eq:Ham2}), we
observe in Figure \ref{graph_3}(b) that the outer invariant curves
(together with the islands) have been destroyed and the dynamical
aperture of the beam has been significantly reduced.

\begin{figure}[ht]
\begin{center}
\includegraphics{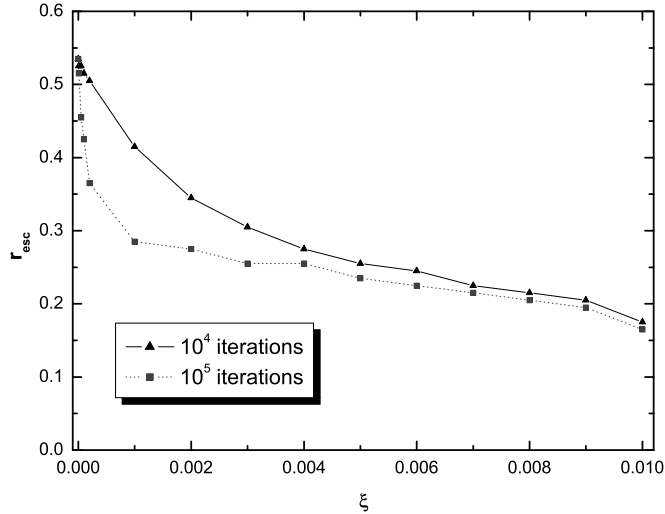}  \vspace{6.5cm}
\end{center}
\caption{Plot of dynamical aperture estimates $r_{esc}$ of a beam
with 2-dimensional cross-section, resulting from $N=10^4$ and
$N=10^5$ iterations of the 4D map (\ref{eq:map2}), with tunes
$q_x=0.61903, q_y=0.4152$ for increasing values of $\xi$ in
(\ref{eq:Ham2}). } \label{graph_4}
\end{figure}

Studying this reduction as a function of $\xi$, we proceed to
choose initial conditions from a grid of step size 0.05, extending
from -0.65 to 0.65 in all 4 directions about the origin,
represented by $X(0),P_1(0),Y(0),P_2(0)$. Iterating the resulting
orbits from points within spherical shells of width $\Delta
r=0.01$, we plot in Figure \ref{graph_4}, for each $\xi$, the
$r_{esc}$ value of the inner radius of the shell from which at
least one orbit escapes to infinity. Our results demonstrate that
the beam's dynamical aperture steadily decreases as $\xi$ grows.
At $\xi=0.006$ (or $\omega_1/\omega_{01} \simeq 0.98,
\omega_2/\omega_{02} \simeq 0.955$) its radius has fallen by more
than 50 percent from its original value, while at higher perveance
values the approximation $\xi<<r_c^2$ no longer applies. In fact,
it is worth noting that the size of the dynamical aperture falls
drastically even for small values of $\xi$, as our calculations
with $N=10^5$ iterations show. For example even at $\xi=0.0002$
(or $\omega_1/\omega_{01} \simeq 0.9993, \omega_2/\omega_{02}
\simeq 0.9985$) our estimate of the dynamical aperture has dropped
from 0.54 to 0.37.

\section{Conclusions}

High intensity effects have long been studied in connection with
the so called beam-beam interaction and were a relevant topic in
the design of many hadron colliders like ISABELLE and the SSC (see
articles in \cite{Beam-Beam,Month_1986,Month_1987}). However, the
effects of high currents on the beam  stability have become
especially crucial only in recent times, when the design and
construction of medium energy high current accelerators has
started.

We have reported in this Letter our results on the possible
importance of space charge effects to the global stability of
intense hadron beams, experiencing the sextupole nonlinearities of
an array magnetic focusing elements, through which the particles
pass $N=10^{5-6}$ times in a typical ``medium term" experiment of
intense beam dynamics. We have used a recently developed
analytical approach \cite{BenTurc} to model the space charges by a
``frozen core" distribution, valid to first order in canonical
perturbation theory. By proposing a simple example of such a
distribution, which leads to explicit and convenient formulas, we
have been able to carry out detailed numerical investigations on
perturbations of 2D and 4D mapping models, describing the dynamics
of flat (horizontal) and elliptic beams (with horizontal and
vertical displacements) respectively.

These charge distributions are in effect periodic modulations of
the tunes (and space advance frequencies) of the motion and are
therefore expected to introduce new resonances, raising the phase
space dimensionality of the dynamics. Thus, outer invariant tori
of the unperturbed case start to disappear and ``island" chains of
higher order resonances far from the origin eventually drift away,
leading to a significant decrease of  the region of bounded
betatron oscillations of the particles about their ideal path
(i.e.~the beam's dynamical aperture, or luminosity).

In our experiments, we have been able to measure this reduction of
the beam's dynamical aperture, for several small values of the
perveance parameter $\xi$, representing the strength of the space
charge distribution. We found that, within the range of validity
of our approximations, the domain of bounded orbits decreases by a
significant percentage and hence space charge effects should be
taken into consideration as they can be important for the long
term survival of the beam. In the flat beam case, we observed a
near total loss of the beam at some $\xi$ value, which is most
likely caused by the onset of a major new resonance introduced by
the space charge modulations. On the other hand, in the more
general case of a beam with 2- dimensional cross section modelled
by a 4- dimensional map, we also discovered a sudden drop in the
dynamical aperture, occurring already at very small tune
depressions.

We, therefore, believe that space charges are important enough to
merit further investigation in mapping models of intense proton
beams \cite{BS06}. The occurrence of new low order resonances
poses, of course, a major threat to the dynamics, if the perveance
parameter is big enough. However, even at small values of this
parameter, weak (Arnol'd) diffusion effects and the slow drift of
high order resonances, may significantly alter the long term
picture of the motion, after a sufficiently great number of
iterations. It would also be useful to compare the one turn map
with the full integration of the space charge effect over one beam
revolution to appreciate the validity limits of our approximation.
Indeed, since the high computation efficiency of the one turn map
is a key issue of this approach, an estimate of the errors in some
reference cases would contribute additional useful information in
realistic applications.

\section{Acknowledgments}
We are particularly grateful to the two referees for their very
valuable comments which helped significantly in improving the
exposition of our results. T. Bountis acknowledges many
interesting discussions on the topics of this paper with Professor
G. Turchetti, Dr. H. Mais, Dr. I. Hoffmann and Dr. C. Benedetti at
a very interesting Accelerator Workshop in Senigallia, in
September 2005. He and Ch. Skokos are thankful to the European
Social Fund (ESF), Operational Program for Educational and
Vocational Training II (EPEAEK II) and particularly the Programs
HERAKLEITOS, and PYTHAGORAS II, for partial support of their
research in physical applications of Nonlinear Dynamics.


\end{document}